\documentclass[prl,showpacs,preprintnumbers,amsmath,amssymb,
superscriptaddress,twocolumn]{revtex4}

\usepackage{color,epsfig}
\usepackage{graphicx}
\definecolor{violet}{rgb}{0.4,0,0.4}
\definecolor{vert}{rgb}{0,0.5,0.0}
\definecolor{navy}{rgb}{0.0,0.0,0.6}
\definecolor{orange}{rgb}{0.8,0.2,0.0}
\definecolor{bleu}{rgb}{0.3,0.0,0.8}

\def\be{\begin{equation}}
\def\fe{\end{equation}}

\font\impe=cmss10

\begin{document}

\title{\bf Rigidity and stability of cold dark solid universe model.}

\author{R.A. Battye},

\affiliation{Jodrell Bank Observatory, School of Physics and Astronomy, 
University of Manchester, Macclesfield, Cheshire SK11 9DL, UK.}

\author{B. Carter}

\affiliation{LuTh, Observatoire de Paris, Meudon, 92195 France.}

\author{E. Chachoua}

\affiliation{LuTh, Observatoire de Paris, Meudon, 92195 France.}

\author{A. Moss}

\affiliation{Jodrell Bank Observatory, School of Physics and Astronomy,
University of Manchester, Macclesfield, Cheshire SK11 9DL, UK.}

\date{20th May, 2005}

\begin{abstract}
The large scale dynamics of the universe appears to be dominated by a
``dark energy'' constituent with negative pressure to density ratio  
$w=P/\rho$,  which could be stable if sufficiently rigid, but not if 
purely fluid. It was suggested by Bucher and Spergel that such a 
cosmological solid might be constituted by a cold (static) distribution 
of cosmic strings with $w=-1/3$, or membranes with the observationally 
more favoured value $w=-2/3$, but it was not shown that the rigidity 
actually would be sufficient for stability.  For cases in 
which the defect lattice is formed from even junctions, it is found
that the rigidity to density ratio will be given by $\mu/\rho=4/15$ in 
both the string and membrane cases, and it is confirmed that this is 
indeed sufficient for stabilisation with respect to sufficiently
small perturbations.

\end{abstract}

\maketitle

\section{Introduction}

It was pointed out by Bucher and Spergel~\cite{BS98} that an underlying
dark energy component consisting of a frozen network of topological defects,
having an approximately uniform (unclustered) density distribution $\rho$ and 
a strongly negative value of its pressure $P$, could account for many of the 
observed features of the universe on a large scale. In particular, the cosmic 
acceleration suggested by measurements of type Ia supernovae.
It was suggested~\cite{BBS99} that, due a symmetry breaking phase 
transition at a cosmological temperature of a few hundred KeV or less, such 
a constituent might be provided by a suitable distribution of membranes 
(domain walls), whose averaged tension would be able to provide a negative 
pressure, provided the positive pressure contribution from kinetic effects 
is not too large, i.e. provided the distribution is effectively cold enough  
to be treated as an approximately static ``frozen'' state.
 
For ordinary Nambu-Goto type cosmic strings, Bucher and Spergel
noted~\cite{BS98,BBS99} that in a static isotropic distribution the pressure 
to density ratio, $w=P/\rho$ (in units with unit light speed) would be 
given by $w=-1/3$. Moreover, for ordinary Dirac type membranes (such as 
simple domain walls) in a cellular lattice the value would be $w=-2/3$, which 
is fairly close to what is considered~\cite{Silk04,BM05} to be 
observationally favoured. 

The optimism of earlier~\cite{BS98,BBS99} and some more 
recent~\cite{Silk04,BM05,C04} discussions of such a proposal has been
challenged from various points of view. One form of objection concerns
conceivable observable consequences ~\cite{Starkman01} derived from 
assumptions that are themselves open to question. However, it would seem 
that the most fundamental problem is the absence, so far, of any convincing 
suggestion as to how an effectively ``hot'', wiggly, disordered structure
-- such as would naturally emerge from a Kibble-type spontaneous symmetry 
breaking -- could be damped in such a way as to ``freeze'' as an 
effectively ``cold'' lattice of straight string or flat wall segments, 
as is required for the pressure to become effectively negative. 
The positive pressure contribution from wiggles and waves will of course 
tend to be removed naturally by standard damping mechanisms, but the
problem is that such processes~\cite{PS,McGraw98} will usually do so by 
removing the segments involved (so that the mean pressure to density ratio 
remains positive), and will not necessarily tend toward a regular lattice.

The present work does not attempt to address this basic problem of finding 
a plausible mechanism for setting up a cold string or membrane system of the 
postulated kind in the first place. Our purpose here is just to provide a 
better understanding of the subsequent mechanical behaviour of such a system 
-- assuming it has already been somehow created -- and in particular to 
investigate conditions under which such a negative pressure system can 
be stable.

For a purely fluid medium, negative pressure would of course entail local 
instability on a short timescale, as symptomised by an imaginary value for 
the sound speed $c_{\rm_S}$ given by
\be c_{\rm_S}^{\, 2}=\frac{ {\rm d} P}{{\rm d\rho}}\, . \label{1}\fe
In response to this implied objection, Bucher and Spergel pointed out 
that a medium constituted of cosmic strings or membranes would behave like 
an elastic solid, rather than a perfect fluid, which can be stable even when 
the pressure is negative, provided the rigidity is sufficient. 

The aim of the present work is to justify this claim, at least for membrane 
systems which contain only even junctions. However our calculation does not 
apply to wall systems with odd junctions, which we expect to be much less 
rigid, so that they are likely to be, typically, too unstable for relevance
as a dominant dark energy constituent.  (If the walls were sufficiently 
massive, due to formation by symmetry breaking  at very high energy, systems 
with odd junctions might nevertheless  have been relevant to the primeval  
structure formation mechanism discussed in refs.~\cite{Kubo92,Kubo00}). Our 
quantitative estimates will be based on the supposition that it is a 
sufficiently good approximation to treat the system as effectively isotropic 
on a macroscopic scale  (meaning large compared with the mesoscopic lattice 
separation scale) but this simplifying condition does not seem essential for 
our qualitative conclusions, which we expect to remain valid even after 
allowance for small deviations from macroscopic anisotropy.

\section{Distinction between ``even'' and ``odd'' type systems}

The calculation we present here will specifically be concerned with membrane 
systems that can be described as even-type, meaning systems where the number 
of walls intersecting at each (string-like) junction is even. Alternatively, 
if the number of walls meeting at each junction is odd, this will described as 
of odd-type. Mixed systems containing both odd- and even-type junctions might 
arise in sufficiently complicated field models, but systems of that kind 
have not been put forward to date and will not be considered here

\begin{figure}
\centering
\epsfig{figure=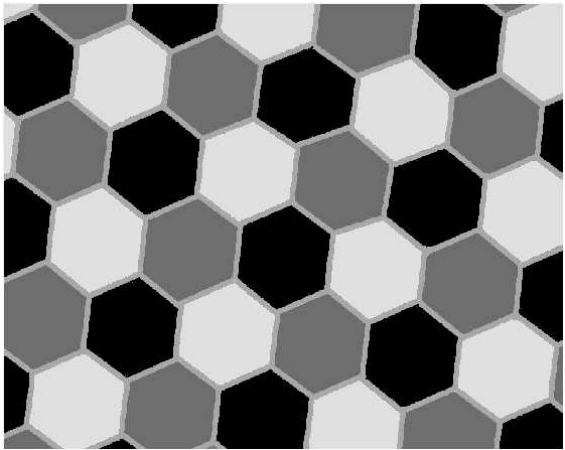, height=6 cm}
\caption{Illustration of two dimensional lattice of the {\it odd} type, to 
which the present analysis does {\it not} apply. Each of the three colours 
corresponds to a different vacuum state. This hexagonal lattice is provided by
a complex scalar field model of the form (\ref{comp}) with $M=3$. 
}\label{fig:hex}\end{figure}

\begin{figure}
\centering
\epsfig{figure=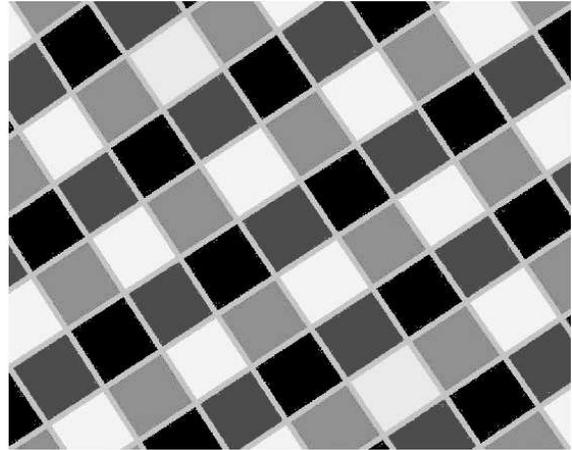, height=6 cm}
\caption{Analogue of Fig.~\ref{fig:hex} with four differently 
coloured vacuum states obtained from (\ref{comp}) by setting $M=4$. 
This case provides a lattice of the {\it even} type to which the present 
analysis applies.
}\label{fig:444}\end{figure}

Simple odd- and even-type systems can be obtained in two dimensions from 
complex scalar field models with a potential 
\be 
V\{\Phi\}=\lambda |\Phi^M-\eta^M|^2\, ,\label{comp}
\fe
where $M$, $\lambda$ and $\eta$ are constants. There are $m$ minima of this 
potential at $\Phi=\eta\exp[im/M]$ for $m=0,1,..,M-1$. For $M=3$ one can 
construct a hexagonal lattice as illustrated in Fig.~\ref{fig:hex} which 
is clearly of the odd-type, whereas for $M=4$ an even lattice as 
illustrated in Fig.~\ref{fig:444} is possible. We note that the stability of 
such systems, which can be derived from a super-potential~\cite{gary}, has 
been investigated in ref.~\cite{saffin}.

Our analysis will be based on the postulate that the affine displacements 
completely describe the modification of the equilibrium distribution at a
mesoscopic level (meaning large compared with the membrane thickness, but 
small compared with the lattice separation scale), that is, we will assume 
that the local equilibrium of the system will preserved by arbitrary affine 
space transformations. We describe this as the {\it affine equilibrium 
preservation postulate.}

\begin{figure}
\centering
\epsfig{figure=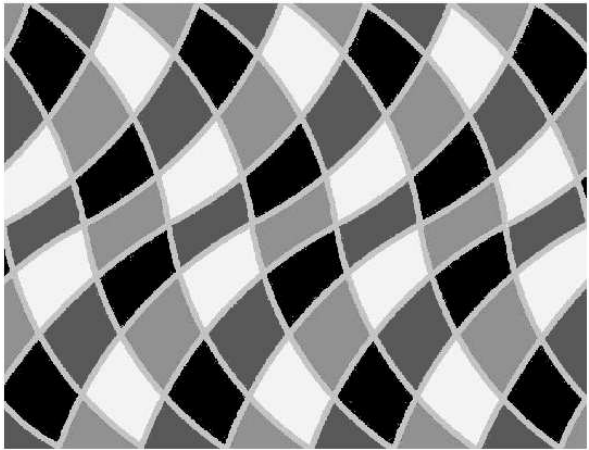, height=6 cm}
\caption{The effect of a (vertically propagating) shear wave of simple
mesoscopically affine type on the even type lattice illustrated in 
Fig.\ref{fig:444}. It is clear that the local equilibrium of each of 
the junctions is maintained under this locally affine deformation. 
This would not be possible for the hexagonal lattice illustrated 
in Fig. \ref{fig:hex}, for which a shear deformation could be affine
only at a macroscopically averaged scale.
}\label{fig:444sh}\end{figure}

This postulate is applicable to lattices of the even-type, for which the flat 
membrane sheets  cross straight through each other (as in an {\impe X}-form) 
at an angle that can freely adjusted without affecting the condition of  
equilibrium. However, it can be expected to fail for odd-type systems:  
for example at a ({\impe Y}-form) junction between 3 equivalent  membrane
segments in equilibrium the intersection angle must always be  $2\pi/3$, a 
condition that will not be preserved by a deformation that is affine on a 
mesoscopic scale. In such a case, a macroscopically affine deformation would
therefor entail non affine (rigidity lowering) adjustments at the mesoscopic
 level. The effect of a locally affine (shear wave) transformation on the two 
dimensional complex scalar field model with $M=4$ is illustrated 
in Fig.~\ref{fig:444sh}

More complicated, and possibly stable, systems of both odd and even type 
type can be obtained from perturbed O(N) models with potential energy 
density given by N real scalar fields $\Phi_i\, ,$ and mass scales 
$\eta\, ,\ \zeta\, ,$  in the form 
\be  V\{\Phi\}=\lambda \,(\sum_i\Phi_i^2-\eta^2)^2+
{\cal E}\, \sum_i (\Phi_i^2-\zeta^2)^2 \,  , \label{ON} \fe
where $\lambda $ and $ {\cal E} $ are dimensionless parameters such that 
$ {\cal E} > -\lambda\, ,\ {\rm N}\lambda >- {\cal E} .$ The exactly  O(N)
symmetric model is obtained by taking ${\cal E}=0\, ,\ \lambda>0$, while 
the simplest example of an even-type system is provided by the decoupled 
limit, $\lambda=0\, ,\ {\cal E}>0$, for which the domain walls simply pass 
through each other without interaction. 

Less trivial possibilities include the systems with $\zeta=0$ considered by 
Kubotani and  collaborators~\cite{Kubo92,Kubo00}, who specifically envisaged 
odd-type systems -- with triply intersecting ({\impe Y}-form) junctions --  
obtained for N=3 by taking ${\cal E}<0$. In this case there are 6 vacua which 
can be thought of as being at the centre of the faces of a cube. The 
alternative, of relevance  for the present work, is the even case -- with  
quadruple  ({\impe X}-form) junctions~\cite{C04} between walls separating 
vacuum  domains where $\Phi_i^2=(\lambda\eta^2+{\cal E}\zeta^2)/({\rm N}
\lambda +{\cal E})$ for $i=1,..,N$ -- which can be obtained (for any value of 
N) by taking the  symmetry breaking parameter to be positive, ${\cal E}>0\,$. 
For N=3 there are 8 vacua at the vertices of a cube.

\begin{figure}
\centering
\epsfig{figure=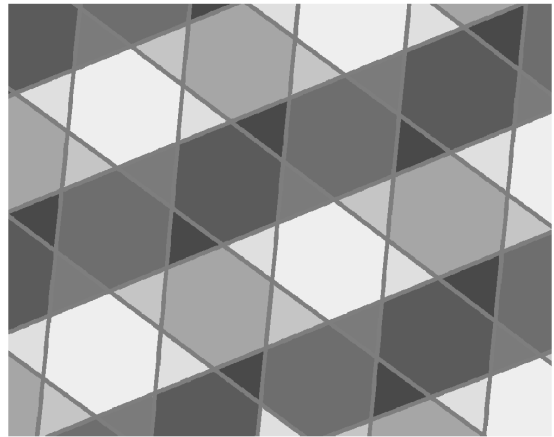, height=6 cm}
\caption{
Illustration of a 2 dimensional section through a frozen cubic lattice of 
``even'' type, having (differently shaded) domains bounded by flat walls 
with {\impe X} form junctions, obtained for system with eightfold vacuum  
provided by broken O(N) model with $\varepsilon>0$ and N=3. Such a
configuration has an averaged stress tensor that will already be exactly 
isotropic, but to get a very highly isotropic elasticity tensor a rather 
larger value of N would be needed.}\label{fig:cubic}\end{figure}

Field models like this engender lattices of periodic or quasi-periodic type, 
with walls grouped in mutually parallel families, the prototype example 
being that of the cubic lattice for N=3, in which there are just 3 families 
arranged so as to be mutually orthogonal. The question of whether disordered  
(glass-type) configurations can be obtained from other plausible models 
remains unclear. Although its averaged stress tensor will be isotropic, the 
cubic (N=3) case (see Fig.~\ref{fig:cubic}) will be characterised  by an 
elasticity tensor that distinguishes between directions parallel to the planes 
of the walls and directions that can deviate from these planes by as much as 
$\pi/4$. For larger values of N, and correspondingly larger numbers of 
families of mutually parallel walls, the possible deviations will get 
progressively smaller so that, like the stress, the ensuing elasticity 
properties will become highly isotropic, as postulated in the analysis below.

\section{Local stability condition}

For a system of the even-type just described, a lattice of flat membrane 
states continuing straight through each other -- as if without interaction 
-- at simple crossover junctions will evidently provide a local equilibrium 
configuration in a flat space background. Such an equilibrium will clearly 
be stable at the mesoscopic scale characterised by the lattice cells, because 
the unbounded -- effectively non-interacting -- membrane sheets are each 
individually stable with respect to local perturbations, provided their 
amplitude is small compared with the lattice spacing, so as to prevent 
distinct vacuum domains of the same type from colliding. 

Assuming that -- at a macroscopic scale large compared with the cell spacing 
-- such a lattice interacts with other (relatively lightweight) constituents 
of the cosmic background sufficiently to form an effectively  coherent 
medium (like a gas of point particles on scales large compared with the mean 
free path) one would expect that this medium would also be stable, which 
means having enough effective rigidity to provide real values for the 
relevant squared velocities of propagation.

To verify that, for an approximately isotropic unperturbed configuration, 
this will indeed be so, we need the formula for the effect of rigidity on 
the relevant propagation velocity. This was first studied ~\cite{C73} when 
relativistic elasticity theory was originally developed for application to 
the solid crust of a neutron star~\cite{C83}, but the same formalism applies 
equally well here. As in the terrestrially familiar non-relativistic limit, 
there will be transversely polarised  -- shake type -- modes with 
propagation speed $c_{_\perp}$, as well as a longitudinally polarised -- 
sound type -- mode with propagation speed $c_{_\Vert}$. 

It was shown~\cite{C73} that the latter would exceed the ordinary fluid 
sound speed value (\ref{1}) by an amount proportional to the rigidity 
modulus $\mu$ according to the formula
\be c_{_\Vert}^{\, 2}=c_{\rm _S}^{\, 2}+\frac{4}{3}\,c_{_\perp}^{\, 2}
\, ,\label{2}\fe
in which the squared shake mode speed is given by
\be c_{_\perp}^{\, 2}=\frac{\mu}{\rho + P}\, .\label{3}\fe

For stabilisation of the medium, that is to ensure that $c_{_\Vert}^{\, 2}$ 
will be positive even if $c_{\rm _S}^{\, 2}$ is not, the necessary and 
sufficient condition is therefore
\be \mu> -\frac{3}{4}\,\beta\, ,\label{4} \fe
where $\beta$ is the bulk modulus, as defined in the relativistic
case~\cite{C73,C83} by the formula
\be \beta=(\rho +P)\,\frac{ {\rm d} P}{{\rm d\rho}}
\, .\label{16} \fe
This reduces to the familiar form $\beta=\rho\, {\rm d}P/{\rm d}\rho$
in the non-relativistic limit $\vert P\vert\ll\rho$. The requirement 
(\ref{4})  simplifies to $ \mu> -3\,\gamma P/4$ in the case of a polytropic 
equation of state,  defined in terms of a conserved number density $n$ and
 constants $\gamma$, $\kappa$, and $m$ by $\rho=\kappa\, n^\gamma+m\, n$ which 
implies $P=(\gamma -1)\,\kappa\, n^\gamma$. 

The models considered here correspond to the zero-mass limit, $m=0$, in this 
polytropic class, and the bulk modulus will be given by
\be \beta= w \,\gamma\,\rho\, ,\hskip 1 cm  w=\gamma-1\, .\label{17}\fe
Since we have $\gamma=2/3$ in the string case, and $\gamma=1/3$ in 
the membrane case, it follows that the bulk modulus will have a negative, 
and thus destabilising, value that will be given  by the same formula, 
\be \beta=-\frac{2}{9}\,\rho\, ,\label{18}\fe
in both string and membrane cases. It can be seen that this result
will still be valid for a non-polytropic mixture with $\rho=\kappa_1\, 
n^{1/3}+\kappa_2\,n^{2/3}$. Therefore, either separately 
and also for a mixed system of strings and membranes, the stability 
criterion (\ref{4}) will reduce to a requirement of the same simple form, 
\be\frac{\mu}{\rho}>\frac{1}{6}\, .\label{19}\fe

\section{Evaluation of rigidity}

The new result provided by the present work for the even-type lattices
described above is the explicit evaluation of the rigidity $\mu$, and the 
confirmation that the ratio $\mu/\rho$ -- which turns out to be the same for 
strings as for membranes -- actually will satisfy the stability 
condition (\ref{19}).

The required rigidity, or shear, modulus $\mu$ can be defined, for an 
initially isotropic state, by expressing the static -- quadratic order -- 
change in the mass-energy density $\rho$ due to an infinitesimal volume 
conserving space displacement as~\cite{LL,C73,C83}
\be \Delta\rho=\mu\, e_{ij}\, e^{ij}\, ,\label{20}\fe
where $e_{ij}$ is the 
infinitesimal strain tensor. For an affine displacement 
given in Cartesian coordinates by 
$x^i\mapsto\tilde x^i=x^i+\varepsilon^i_{\,j}x^j$, it is 
defined as the symmetric part of the deformation matrix: 
$e_{ij}=(\varepsilon_{ij}+\varepsilon_{ji})/2 $. 

In order to avoid an extra energy variation contribution due to the pressure, 
we impose the restriction that the displacement must conserve volume. This 
requires that the Jacobean determinant $\vert\partial\tilde x_i/
\partial x_j\vert$ should be unity which implies that the trace $e^i_{\, i}$, 
like the variation $\Delta\rho$, will vanish to linear -- though not higher 
-- order in $\varepsilon$. An obvious way to do this is to take the symmetric 
(zero curl) pure shear deformation given in terms of coordinates 
$x^{_1}=x$, $x^{_2}=y$, $x^{_3}=z$ by $\tilde x=(1+\varepsilon)x$, 
$\tilde y=y$, $\tilde z=z/(1+\varepsilon)$, which when substituted 
in (\ref{20}) gives 
\be \Delta\rho=2\,\mu\ \varepsilon^2\,,\label{5}\fe 
to second order in $\varepsilon$.  The value of $\mu$ can be read out from 
this when the variation $\Delta\rho$ has been worked 
out. 

However, the required value of $\mu$ can be obtained in a 
computationally more convenient way by considering the effect on $\rho$ of 
the  asymmetric (non-zero curl) simple shear deformation given by
\be \tilde x=x+2\,\varepsilon\, z\, ,\hskip 1cm \tilde y=y\, ,\hskip 1 cm
\tilde z=z\, ,\label{22}\fe
which is of the kind produced dynamically by a transverse mode with 
propagation speed in the $z$ direction, given by (\ref{3}). It can be 
checked that a simple shear deformation such as this will provide a second 
order expression of the same form (\ref{5}) to the one obtained for the pure 
shear deformation considered above. 

To evaluate the density variation $\Delta\rho\, $ produced -- as the left
hand side of (\ref{5}) --  by the action of the simple volume conserving 
transformation (\ref{22}) on a distribution of randomly oriented strings, 
let us consider the case of a sample string segment from the origin to a 
point with coordinates given by the components,  $\{x,y,1\}$ say, of a 
vector $\vec\ell$,  which will be mapped to a vector 
$\vec\ell+\Delta\vec\ell$ with coordinates $\{x+2\,\varepsilon, y, 1\}$. 
The length $\ell=(x^2+y^2+1)^{1/2}$ of the segment will undergo a 
fractional change given, to quadratic order in $\varepsilon$, by
\be\frac{\Delta\ell}{\ell}=\frac{2\,x\,\varepsilon}{\ell^2}+
\frac{2(\ell^2-x^2)\varepsilon^2}{\ell^4}\, .\label{6}\fe
Since the mass-energy of a static Nambu Goto string segment is obtained 
simply by multiplying its tension, $T$ say, the formula (\ref{6}) 
immediately provides the fractional change in the contribution to the mass 
density $\rho$ from strings with the chosen direction, as given in polar 
coordinates by $x={\rm tan}\,\theta\, {\rm cos}\,\phi$,
$y={\rm tan}\,\theta\, {\rm sin}\,\phi$.

Since the model is based on the postulate that the strings have a random 
isotropic distribution, their net effect can be computed by just taking the 
spherical average of (\ref{6}) over the polar coordinate values $\theta$, 
$\phi$, which gives a formula in which the first order part cancels out, 
leaving just the quadratic term,
\be\frac {\Delta\rho}{\rho}=\left\langle\frac{\Delta\ell}{\ell}
\right\rangle=\frac{8}{15}\,\varepsilon^2\, .\label{7}\fe

To evaluate the effect of the same simple shear deformation (\ref{22}) on 
an isotropic  distribution of plane membranes, consider a parallelogram 
shaped sample segment spanned by two vectors, $\vec \ell_{_1}$, with 
components $\{x_{_1},1,0\}$, which will not change, and $\vec\ell_{_2}$, 
with components  $\{x_{_2},0,1\}$, which will be deformed to $\{(x_{_2}
+2\varepsilon),0,1\}$. The membrane analogue of the string formula 
(\ref{7}) for the effect on the density of an isotropic distribution is
\be\frac {\Delta\rho}{\rho}=\left\langle\frac{\Delta A}{ A}
\right\rangle\, ,\label{8}\fe
where $A$ is the sample area. This is given by the magnitude of the vector 
product $\vec A=\vec\ell_{_1}\times \vec\ell_{_2}$, with components 
$\{1,x_{_1}, -x_{_2}\}$ in the direction normal  to the membrane. The 
deformation (\ref{22}) will map this area-calibrated normal vector to 
$\vec A+ \Delta\vec A$, with components $\{1,x_{_1}, -(x_{_2}+2\,
\varepsilon)\}$. Since this transformation differs from that of the string 
segment considered above only by a permutation of the roles of the $x$ and 
$z$ axes, the final result, after spherical averaging,  will be the same. 

For the macroscopically isotropic and mesoscopically even-type systems 
under consideration, this reasoning provides a theorem to the effect 
that -- like the bulk modulus $\beta$ as given by (\ref{18}) -- the 
rigidity modulus $\mu$ of the isotropic cosmic membrane distribution will 
be identical to that of an isotropic cosmic string distribution with the 
same density $\rho$. Thus, by comparing (\ref{7}) with (\ref{5}), it can be 
seen that the value of the rigidity is 
\be \mu=\frac{4}{15}\,\rho\, ,\label{9}\fe
not only in the string case, but also in the membrane case and for a 
mixed system of strings and membranes.

\section{Conclusions}

It is evident that the value given by (\ref{9}) does indeed satisfy the 
stability condition (\ref{19}). This 
confirms the intrinsic coherence of the solid universe model subject to the 
availability of an underlying field theory of a suitable kind, meaning one 
providing a membrane system of even-type, for example, (\ref{ON}) 
with ${\cal E}>0$. However, it remains unclear how such a system could have 
come into existence in the first place.

In a solid of this kind, as a consequence of (\ref{2}) and (\ref{3}), we 
obtain real values for the speeds of longitudinal as well as transverse 
modes. For the latter, (\ref{3}) with (\ref{9}) gives 
$ c_{_\perp}^{\, 2}={4}/{15\gamma}$, so the (transverse) shake wave speed 
will be given by
$$ \gamma=2/3\ \Rightarrow\  c_{_\perp}^{\, 2}=2/5\, ,\hskip 0.6 cm
\gamma=1/3\ \Rightarrow\  c_{_\perp}^{\, 2}= 4/5\, ,\label{11}$$
for the string and membrane cases respectively. The corresponding formula 
obtained from (\ref{2}) for the longitudinal wave speed will take the form 
\be c_{_\Vert}^{\, 2}=w+\frac{16}{45\gamma}\, ,\label{12}\fe
which will indeed be consistent with the stability criterion
(\ref{4}),  since it clearly provides a real-valued propagation speed 
for any positive value of the polytropic index,
$$ \gamma=w+1 \, .$$ 
In particular it can be seen that the longitudinal propagation speed will 
be given  by 
$$ w=-1/3\  \Rightarrow\  c_{_\Vert}^{\, 2}=1/5\, ,\hskip 0.6 cm
w=-2/3\ \Rightarrow\  c_{_\Vert}^{\, 2}=2/5\, ,$$
in the string and membrane cases respectively, the latter being the one 
of greatest cosmological interest.

The present analysis is based on the cold limit in which thermal~\cite{C94} 
or other kinetic excitations of the world sheets are neglected.  If 
significant, such effects would tend~\cite{C95} to reduce the effective 
tension, giving slower perturbation speeds and also making it harder to 
explain the apparent~\cite{Silk04} cosmic acceleration. Such effects
would also tend to undermine the stability of the system, which is valid 
(as shown above) only for perturbation amplitudes that are small compared 
with the relevant lattice separation scale. The fundamental difficulty 
with this kind of scenario is the problem of finding a natural mechanism 
for getting rid of such excitation effects so as to actually obtain a frozen 
lattice of the required kind.

Assuming that the membrane domination scenario can be achieved, and that it 
could be obtained by some spontaneous symmetry breaking process, a first 
attempt was made~\cite{BBS99} at evaluating the energy scale, $\eta$ say, 
characterising the tension, $T\approx \eta^3$, of the required membranes and 
the cosmological temperature, $\Theta\approx \eta$, of their presumed 
formation. It relied on the assumption that the relevant correlation 
lengthscale at the epoch of formation would be the maximum permitted by 
causality. In a recent reconsideration of this issue by one of the present 
authors~\cite{C04} it has be shown that, although it would be more realistic 
to suppose the correlation length to be very much smaller than the causal 
limit, nevertheless the effect of the appropriate correction on the estimate 
for $\eta$ is relatively moderate: the likely range is merely reduced from 
the order of a hundred KeV to a few KeV. This does not change the 
qualitatively important consequence that the transition should have occurred 
at a relatively recent cosmological era, after pair annihilation and Big-Bang 
Nucleosynthesis, though before the epoch of recombination when the universe 
became transparent. The corollary is that the present lattice separation 
scale of the membrane lattice will be comparable~\cite{BBS99,C04}, at most, 
with interstellar rather than intergalactic separation lengthscales. Thus, on 
the presumption that the relevant fields interact very weakly with ordinary 
particles such as electrons, and particularly photons, so that the walls will 
not be directly visible, Bucher and Spergel's provisional conclusion that 
their scenario poses no hierarchy problem remains valid, in the sense that 
the range of conceivable values for $\eta$ overlaps with the mass range for 
already known elementary particles. 

Despite this attractive feature, the plausibility of such a scenario is 
undermined by the absence, as remarked above, of any mechanism for getting 
from a randomly excited initial state to a regular frozen membrane lattice 
of the required kind, as exemplified by the case depicted in Figure 1. In view 
of the positive results of our study, this issue warrants further 
investigation.

The authors wish to thank Martin Bucher for instructive conversations.

\end{document}